\documentclass[preprintnumbers,amsmath,amssymb,nofootinbib
,superscriptaddress]{revtex4}

\usepackage{hyperref}
\usepackage{graphicx}
\usepackage{color}
\usepackage{xcolor}
\usepackage{comment}
\usepackage{lmodern}
\usepackage{amssymb}
\usepackage{lipsum}
\usepackage{mwe}

\newcommand{\backo}{\!\!\!\!\!\!\!\!\!\!}
\newcommand{\ebo}{\eea }
\newcommand{\bbo}{\bea  && }

\newcommand{\bl}{\biggl(}
\newcommand{\br}{\biggr)}
\newcommand{\Tr}{\makebox{Tr}}
\newcommand{\vvx}{\vec{x}}
\newcommand{\vvy}{\vec{y}}
\newcommand{\vvr}{\vec{r}}

\newcommand{\be}{\begin{equation}}  
\newcommand{\ee}{\end{equation}}  
\newcommand{\bea}{\begin{eqnarray}}   
\newcommand{\eea}{\end{eqnarray}}  
\newcommand{\ba}{\begin{array}}  
\newcommand{\ea}{\end{array}}

\usepackage{diagbox}

\newskip\humongous \humongous=0pt plus 1000pt minus 1000pt

\newif\ifdtup

\def\oldreffmt#1{\rlap{[#1]} \hbox to 2\parindent{}}

\def\figfmt#1{\rlap{Figure {#1}} \hbox to 1in{}}  
  
\def\etal{\hbox{\it et al.}}  
  
\def\Tr{\mathop{\rm Tr}}

\def\slash#1{#1\!\!\!/\!\,\,} 	
\def\beq{\begin{equation}}  
\def\eeq{\end{equation}}  
\def\bea{\begin{eqnarray}}  
\def\eea{\end{eqnarray}}  
\def\half{\frac{1}{2}}  
  
\def\bq{\begin{quote}}  
\def\eq{\end{quote}}

\def\half{\frac{1}{2}}    
\newcommand{\nonumbo}{ \nonumber \\ && }

\def \etal {{\it et al.}\ }  
\relax

\newdimen\tdim  
\tdim=\unitlength  
\def\bar{\overline}

\RequirePackage{caption}
\begin{document}

\preprint{Delafield-0820-2025}

\title{ Quantum Aspects of 
Natural Top Quark Condensation
}
\author{Christopher T. Hill}
\email{chill35@wisc.edu}
\affiliation{Fermi National Accelerator Laboratory,
P. O. Box 500, Batavia, IL 60510, USA}
\affiliation{Department of Physics, University of Wisconsin-Madison, Madison, WI, 53706
}

\begin{abstract}
 
 In top quark condensation
the Brout-Englert-Higgs (BEH) boson is a $\bar{t}t$ bound state. 
With  a  UV completion of a single coloron exchange interaction 
a recent semiclassical treatment, \cite{main,main2}, gave a novel theory of the BEH boson as an extended
object with composite scale  $M_0\sim 6$ TeV. 
Presently we obtain the semiclassical theory
as an effective action using the source/Legendre transformation techniques 
of Jackiw, {\etal}$\!\!,$ \cite{Jackiw,CJT}, and fermion loop effects  
in the  large-$N_c$ limit by deploying an auxiliary field to implement the sum of leading fermion loop 
diagrams.  The theory remains natural at the loop level with fine tuning at the level
of a few \%, and the effective coupling of the 4-fermion
interaction, $\bar{g}_0^2$, is significantly
enhanced by quantum loops over the fundamental coloron coupling, $g_0^2$. Hence a relatively weaker 
``topcolor'' theory can produce critical coupling in the effective BEH bound state theory.

\end{abstract}

\maketitle
 
\date{\today}


\email{chill35@wisc.edu}


\section{Introduction }

Recently  we revisited  the idea of ``top quark condensation,'' 
i.e., that the  Brout-Englert-Higgs (BEH) boson is composed of top + anti-top quarks \cite{main,main2}.
Our new approach describes a bound state, consisting of a pair of relativisitic chiral fermions, arising from
  a single particle exchange of a perturbative massive  gluon-like object 
  with mass $M_0$ and coupling $g_0$  (dubbed a ``coloron'').  A semiclassical analysis, involving
``bilocal fields,'' leads to a low-mass composite 
BEH boson with an extended internal 
wave-function, $\phi(r)$ \cite{main,main2}.
$\phi(r)$ satisfies a Schr\"odinger-Klein-Gordon (SKG) equation in a Yukawa potential with coupling $g_0^2$. 
Close to its critical coupling, $g_0^2 \sim g_c^2\approx 8\pi^2/N_{color}$ 
the internal wave-function  spreads, thus diluting $\phi(0)$.
The BEH boson is then an extended object with a small eigenvalue, $\mu^2$.
We identify $\mu^2=-(88)^2$ GeV$^2$, the BEH mass in the symmetric phase of the
standard model (SM).
The composite
scale, the mass of the $SU(3)$ octet colorons, 
is then found to be $M_0\sim 6 $ TeV.
Near critical coupling the low energy theory is approximately scale invariant, in analogy to a second order phase transition in condensed matter physics. This effective scale invariance leads to the spreading of the wave-function $\phi(r)$, and is a universal phenomenon in any bound state system near its critical coupling, where the eigenvalue of the Hamiltonian is small compared to the internal parameters of the theory. Hence scale symmetry is realized dynamically as the custodial symmetry of the small mass BEH bound state.

Remarkably, the static observables, 
such as the induced top--quark BEH--Yukawa coupling, $g_Y$, and 
quartic coupling, $\lambda$, depend upon $\phi(0)\sim \sqrt{|\mu|/M_0}$ and become concordant with experiment \cite{main}. 
Moreover, due to a subtlety of the  dilution
effect, the theory requires fine--tuning at
the level of  $\sim |\phi(0)|^2 \sim |\mu|/M_0$,
rather than the usual $|\mu|^2/M_0^2$. Hence, identifying $\mu^2\sim -(88)^2$ GeV$^2$
we require fine tuning only at the level of $\sim$ few $\%$.
 The colorons may be accessible to the LHC, where they would be pair produced and 
 decay to $\bar{t}t$ pairs (also $\bar{b}b$ in extended topcolor models \cite{Topcolor,Topcolor2}) 
 in boosted jets.
 This is significantly different than old results obtained
in the 1990{}'s \cite{Nambu1}$-$\cite{NSD} which were based upon the renormalization group (RG) improved
Nambu--Jona-Lasinio model (NJL, \cite{NJL}), but
did not contain an internal wave-function $\phi(r)$.

Presently we show how the semiclassical theory
emerges from the underlying quantum field theory of the third generation quarks
with the single coloron exchange interaction.
We utilize the formalism developed in the early papers, mainly 
of Jackiw, {\etal}$\!\!,$ \cite{Jackiw}\cite{CJT}, which yield
the effective semiclassical action.
We also introduce auxiliary fields that facilitate the 
sum of the leading fermion loop diagrams of order $(\hbar g_0^2 N_c)^n$.
The approach can be considered a decomposition of a top quark field into
semiclassical + quantum degrees of freedom,
$\sim \psi + \sqrt{\hbar}\omega$.
Integrating out the quantum fermionic fluctuations, $\omega$, in the presence of
fermionic source fields, $\xi_{L,R}$,  yields 
a semiclassical effective ``source'' action. The action for the semiclassical fields, $\psi_{L,R}(x)$,
and a  composite bilocal Brout-Englert-Higgs field, $H_i(x,y)$,
is then generated upon performing a
Legendre transformation to remove the sources.
The resulting effective action then allows treatment of
all semiclassical phenomena in terms of the c-number, chiral spinor fields, $\psi_{L,R}(x)$,
and formation of the bound state $H_i(x,y)$, reproducing our earlier work
(reference \cite{main} should be considered a companion to the present paper).

Mainly,
we find that the sum of $\omega$-loops with coloron coupling, $g_0$, generates a  renormalized coupling
for the  4-fermion interaction,
$\bar{g}^2_0$, where:
\bbo
\bar{g}_0^2 = g_0^2\left(1-\frac{g_0^2N_c}{8\pi^2} \right)^{-1}
\ebo
Essentially, the fermion loops act as in the  NJL model, which 
enhances the binding of the $0^+$ channel ground state, leading to a form of ``critical amplification.''
The underlying coloron production, scattering and decay amplitudes are therefore governed by $g_0^2$, while
the binding, which forms the BEH boson, is governed by the enhanced $\bar{g}_0^2>g_0^2$.
This implies that the $\bar{g}_0^2$ can be  supercritical, while 
the underlying coloron coupling $g_0^2$ remains smaller and subcritical. 
Therefore the effective potential coupling in the $0^+$ channel, $\bar{g}_0^2$, can be 
super-critical, to produce the low mass  BEH bound state and the 
spontaneous symmetry breaking, in a weaker underlying topcolor model. 

The induced top-quark BEH-Yukawa coupling $\bar{g}_Y\propto \bar{g}^2_0$ is also renormalized. Using the experimental
value as an input, $\bar{g}_Y\approx 1$,
we still obtain the composite scale (the coloron mass) $M_0\approx 6$ TeV. 
We also observe how the BEH quartic coupling, ${\lambda}\propto \bar{g}_0^8 \propto \bar{g}_Y^4$, is generated by the underlying $\omega$-loops. In our 
present scheme we find numerically that this is 
remarkably close to the SM result $\lambda\sim 0.25$, which represents a significant improvement
over  the old NJL model based top condensation that 
yielded $\lambda\sim 1$.

This also resolves putative issues that arise in the treatment of ref.\cite{main}.
One might be tempted to, e.g., 
loop the Yukawa  interaction to argue that there is a problematic large correction to the BEH mass $\propto -\bar{g}_Y^2 N_c M_0^2/8\pi^2$.
This would be an additive correction to $\mu^2$
and one might conclude that the effective theory is therefore ``unnatural.''
{\em This would, however, be incorrect.}  This particular
contribution is now already generated by integrating out the underlying $\omega$-fields
and it would be double counting to do a fermion loop calculation in the effective semiclassical theory. 
Moreover, in the underlying theory this  appears, not as a mass term, but as a 
multiplicative and enhancing correction to the effective single coloron exchange interaction and
the internal wave-function back-reacts to it.
It is, in fact, the source of the critical amplification  of $\bar{g}_0^2$.

We find the Jackiw,  {\etal}$\!\!,$ techniques to be powerful, and work particularly well with the auxiliary field fermion
loop sum technique introduced here.

\section{Effective Semiclassical Action}

\subsection{Sources and Legendre Transformation}
 
We presently derive the effective top quark condensation action 
for a semiclassical chiral fermion fields $\psi_{iL}(x)$
and $\psi_R(x)$,
from the underlying quantum field theory, by integrating out third generation, color triplet, quark fields, 
$(\omega_{iL}(x), \omega_R(y))$, where:\footnote{Note that
the $b_R$ quark must also participate in topcolor in an anomaly-free scheme,
which can be implemented as in {\em e.g.}, \cite{Topcolor}.
}
 \bbo
 \label{oneone}
 \omega_{iL}= \frac{(1- \gamma^5)}{2}\bl\begin{array}{c} t\\ b \end{array}\br, \qquad \omega_R = \frac{(1- \gamma^5)}{2}t,
\ebo
where $i$ is an electroweak $SU(2)$ index.
We will use the methods of Jackiw, \cite{Jackiw}
introducing 
chiral source fields, $\Xi^i_{R}$ and $\Xi_{L}$, then performing a Legendre transformation
to obtain the effective action for the semiclassical fields $\psi_{L,R}(x)$.
We then follow with the method of Cornwall, Jackiw and Tomboulis, (CJT \cite{CJT}),
{\em in the semiclassical theory}
and introduce the two-body bound state wave-function source $Q^i(x,y)$, together with semiclassical fields.
A Legendre transformation then yields the full theory with effective fields $\psi_{L,R}$
and  bilocal BEH boson $H_i(x,y)$.
We will deploy a bilocal auxiliary field, $B(x,y)$, to expedite the sum of fermion loops in
a large $N_c$ approximation.

Our starting point is the fundamental action at the quantum field theory level of the topcolor model
with the single coloron exchange interaction, suitably Fierz rearranged (as given in eq.(28) of ref.\cite{main}):
\bbo\label{qaa}
\backo\backo
S(\omega)
=\!\int\! d^4x \bl [\bar{\omega}^i_L(x)\;i\slash{D}\omega_{iL}(x)]+ [\bar{\omega}_R(x)\;i\slash{D}\omega_{R}(x)]
+g_0^2\!\int\! d^4x \; d^4y \; [\bar{\omega}^i_L(x)\omega_{R}(y)]D_F(x-y)[\bar{\omega}_{R}(y)\omega_{iL}(x)],
\ebo
where $[...]$ denotes color singlet contraction of indices.
The  coloron propagator in Feynman gauge is:
\bbo
iD_{\mu\nu}(x-y)= g_{\mu\nu} D_F(x-y);
\qquad
D_F(x-y) =- \int\frac{1}{q^2-M_0^2}e^{iq(x-y)}\frac{d^4q}{(2\pi)^4}.
\ebo
The interaction can be factorized by introducing a bilocal, isodoublet,  auxiliary field $B^i(x,y)$
of mass dimension $3$:  
\bbo
\label{qb}
\backo
S(\omega,B)
=\!\int\! d^4x \bl  
 [\bar{\omega}^i_L(x)\;i\slash{D}\omega_{iL}(x)]+ [\bar{\omega}_R(x)\;i\slash{D}\omega_{R}(x)]\br
\nonumbo
\backo
+\!\int\! d^4x \;d^4y\;\bl
 g_0 D_F(x-y)\left( B^{i\dagger}(x,y)[\bar{\omega}_{R}(y)\omega_{iL}(x)]+  [\bar{\omega}^i_{L}(x)\omega_R(y)]B_i(x,y) \right)
-[B^i{}^\dagger(x,y) D_F(x-y)B_i(x,y)]\br.
\ebo
The ``equation of motion'' of $B_i$ is then:
\bbo
\label{qc}
B_i(x,y)=g_0[\bar{\omega}_R(y)\omega_{iL}(x)]; \;\;\qquad B^{i\dagger}(x,y)=g_0[\bar{\omega}^i_L(x)\omega_{R}(y)].
\ebo
Substituting the solution of 
eq.(\ref{qc}) back into  eq.(\ref{qb}) yields  eq.(\ref{qaa}).
This is analogous to the treatment of the local NJL model (e.g., as in eqs.(4) and (6) in ref.\cite{main}),
but differs presently by the bilocal factor, arising from a bilocal coloron exchange amplitude, of $D_F(x-y)$. Unlike the NJL model, we do not
treat $B_i$ as a  composite field and we will
integrate it out to sum fermion loops.

We can define the fermion kinetic terms and $B_i$ interaction terms of eq.(\ref{qb}) as a 
matrix $K(x,y)$:
\bbo
\label{qd}
S(\Omega,B)
=\!\int\! d^4x \;d^4y\;\bl  
 [\bar{\Omega}(x)K(x,y)\Omega(y)]
-[B^i{}^\dagger(x,y) D_F(x-y)B_i(x,y)]\br,
\ebo
where, 
\bbo
\label{7}
{\Omega} \equiv \begin{pmatrix} \omega_{iL} \\ \omega_R \end{pmatrix},\qquad 
\bar{\Omega} \equiv \begin{pmatrix} \bar\omega^i_L & \bar\omega_R \end{pmatrix},
\qquad
K(x,y)=\begin{pmatrix} S_F^{-1}(x,y)\;\; & g_0D_F(x-y) B^\dagger(x,y)  \\ g_0 D_F(x-y)B(x,y)  & S_F^{-1}(x,y)  \end{pmatrix}.
\ebo
Here $S_F^{-1}(x,y)$ is the formal inverse of the Feynman propagator,
 $S_F(x,y) = (i\slash{D})^{-1}\delta^4(x-y) $. Note that $K$ is Dirac Hermitian, $\bar{K}=\gamma_0K^\dagger \gamma_0=K$.

We follow Jackiw \cite{Jackiw} and add chiral source terms, $\Xi(x)$, for the fermion fields,
\bbo
\label{qbb}
\backo
S(\Xi, \Omega,B) = \!\int\! d^4x \;d^4y\;\bl  
 [\bar{\Omega}(x)K(x,y)\Omega(y)]
-[B^i{}^\dagger(x,y) D_F(x-y)B_i(x,y)]\br
+ \!\int\! d^4x \bl\bar\Xi(x)\Omega(x)+ \bar\Omega(x)\Xi(x)\br,
\nonumbo
\qquad\qquad\qquad\qquad\qquad\qquad
\makebox{where:}  
\qquad
\qquad
{\Xi} \equiv \begin{pmatrix} \xi_{iR} \\ \xi_L \end{pmatrix}, \qquad 
\bar{\Xi} \equiv \begin{pmatrix} \bar\xi^i_R & \bar\xi_L \end{pmatrix},
\ebo
(note that the sources have opposite chirality to their 
corresponding partners, i.e., $\omega_{iL}\leftrightarrow \xi_{iR} $
and
$\omega_{R}\leftrightarrow \xi_{L} $).
The effective action, $W$, is defined by the path integral:
\bbo
W= -i\hbar \ln \!\!\int\!\! D\Omega D\bar\Omega \; \exp \bl\frac{i}{\hbar}  S(\Xi, \Omega,B)  \br.
\ebo
To evaluate $W$
we begin by shifting the integration variables to cancel cross-terms,  
\bbo
\label{shift}
\Omega(x)\rightarrow \Omega(x) - \int\!\!d^4y\; K^{-1}(x,y) \Xi(y);
\qquad \bar\Omega(x)\rightarrow \bar\Omega(x) - \int\!\!d^4y\; \bar\Xi(y) K^{-1}(y,x) ,
\ebo
and the shifted action becomes:
\bbo
\label{qbbb}
S(\Xi, \Omega,B)
= \!\int\! d^4x \;d^4y\; \bl[\bar{\Omega}(x)K(x,y)\Omega(y)]-\bar\Xi(x) K^{-1}(x,y)\Xi(y)  
-[B^i{}^\dagger(x,y) D_F(x-y)B_i(x,y)]\br.
\ebo
The path integral is now $W=W(\Omega)+W(\Xi)+W(B)$ where,
\bbo
\label{133}
W(\Omega)= -i\hbar \ln \!\!\int \! D\Omega D\bar\Omega \; \exp \bl\frac{i}{\hbar}\!\int\! d^4x \;d^4y\; \bl[\bar{\Omega}(x)K(x,y)\Omega(y)]  \br,
\nonumbo
W(\Xi)=-\!\int\! d^4x \;d^4y\; \bar\Xi(x) K^{-1}(x,y)\Xi(y),
\qquad
W(B) = -\!\int\! d^4x \;d^4y\; \bl
[B^i{}^\dagger(x,y) D_F(x-y)B_i(x,y)]\br.
\ebo
where the quantum loops are contained in $W(\Omega)$.
We then define the effective semiclassical field, $\Psi$:
\bbo
\label{14}
{\Psi} \equiv \begin{pmatrix} \psi_{iL} \\ \psi_R \end{pmatrix},\qquad 
\bar{\Psi} \equiv \begin{pmatrix} \bar\psi^i_L & \bar\psi_R \end{pmatrix}
\ebo
where,
\bbo
\label{15}
\bar\Psi(x) = \frac{\delta W(\Xi)}{\delta \Xi(x)}= -\!\int\!\! d^4y\; \bar\Xi(y) K^{-1}(y,x);
\qquad 
\Psi(x) = \frac{\delta W(\Xi)}{\delta \bar\Xi(x)}= -\!\int\!\! d^4y\;  K^{-1}(x,y)\Xi(y).
\ebo
Hence, inverting these relations,
\bbo
\label{16}
\Xi(x) =  -\!\int\!\! d^4y\;  K(x,y) \Psi(y) ; 
\qquad \bar\Xi(x)= -\!\int\!\! d^4y\; \bar\Psi(y) K(y,x),
\ebo
$\Xi$ should be regarded as a functional of $\Psi$ by eqs.(\ref{16}) .

We then perform a Legendre transformation to obtain  the 
effective action, $S_{}(\Psi, B)\equiv W$, swapping $\Xi$ for the semiclassical field, $\psi$, retaining the auxiliary field $B$:
\bbo
\label{newa}
S_{}(\Psi, B) = W(\Omega)+W(B)+W(\Xi) -\int\! d^4x  \;\left(\frac{\delta W(\Xi)}{\delta \Xi(x)}\Xi(x) + \bar\Xi(x) \frac{\delta W(Xi)}{\delta \bar\Xi}\right)
\nonumbo
 = W(\Omega)+W(B)-\!\int\! d^4x \;d^4y\; \bar\Xi(x) K^{-1}(x,y)\Xi(y) - \int\! d^4x  \;(\bar\Psi(x)\Xi(x) + \bar\Xi(x) \Psi(x))
\nonumbo
=W(\Omega)+\int\! d^4x \; d^4y \; \bl\bar\Psi(x) K(x,y )\Psi(y)
-[B^i{}^\dagger(x,y) D_F(x-y)B_i(x,y)]\br.
\ebo
We obtain the normal-sign kinetic terms and chiralities for the semiclassical fields $\psi_{L,R}$.
From eq.(\ref{7}) we have:
\bbo
\label{resultaa}
S_{}(\Psi, B)= W(\Omega)
+\!\int\! d^4x \bl  
 [\bar{\psi}_L(x)i\slash{D}\psi_{L}(x)]+ [\bar{\psi}_R(x)\;i\slash{D}\psi_{R}(x)]\br
 -\int\! d^4x \; d^4y \; \bl [B^i{}^\dagger(x,y) D_F(x-y)B_i(x,y)]\br
\nonumbo
\qquad \qquad
+\!\int\! d^4x \;d^4y\;
\left( g_0 D_F(x-y)( B^{i\dagger}(x,y)[\bar{\psi}_{R}(y)\psi_{iL}(x)]+  [\bar{\psi}^i_{L}(x)\psi_R(y)]B_i(x,y) \right).
\ebo
The ``equation of motion'' of $B_i$ is now:
\bbo
\label{qc2}
B_i(x,y)=g_0[\bar{\Psi}_R(y)\Psi_{iL}(x)]+\frac{\delta W(\Omega)}{\delta B^{\dagger i}(x,y)},  \qquad 
B^{i\dagger}(x,y)=g_0[\bar{\Psi}^i_L(x)\Psi_{R}(y)]+\frac{\delta W(\Omega)}{\delta B_i(x,y)}.
\ebo
If we ignore the quantum $W(\Omega)$ terms, and insert $B_i(x,y)$ back into the new action, 
we recover the original 
classical action of  eq.(\ref{qaa}), but now given in terms of the semiclassical fields $\Psi$: 
\bbo\label{qclassical}
\backo\backo
S(\Psi)
=\!\int\! d^4x \bl [\bar{\psi}_L(x)i\slash{D}\psi_{L}(x)]+ [\bar{\psi}_R(x)\;i\slash{D}\psi_{R}(x)]
+g_0^2\!\int\! d^4x \; d^4y \; [\bar{\psi}^i_L(x)\psi_{R}(y)]D_F(x-y)[\bar{\psi}_{R}(y)\psi_{iL}(x)].
\ebo
The quantum loop effects are contained in $W(\Omega)$, which we calculate in Section \ref{quantum} .
We have obtained the semiclassical action for the c-number fields $(\psi_{iL}(x),{\psi}_{R}(y))$.

\subsection{Full Effective Action for Bilocal BEH boson via Cornwall-Jackiw-Tomboulis \cite{CJT}}

With the  semiclassical theory we now 
 include a source for a bilocal field  which will
lead to the semiclassical BEH boson field,  $H_i(x,y)$.
Note that, since we have already integrated out  the underlying quantum fields,  $\omega$,
the bound state $H_i(x,y)$ that we obtain presently
{\em is composed only of the semiclassical modes $\psi_{L,R}$}. 
We can operate at the semiclassical level, following 
the formalism of Cornwall, Jackiw, and Tomboulis, (CJT) \cite{CJT}.
 This is distinct from the NJL model, or the usual applications of CJT,  where the bound state is composed
in loops of a sum over all of the $\omega$ quantum fields.

We add an additional 
source term for a complex bilocal field, $Q_i(x,y)$, that couples only {\em to a subset of the semiclassical
 (to-be-bound) modes}, $ [\bar{\psi}_{R}(x)\psi_{iL}(y)]_b$, in  eqs.(\ref{resultaa}), while the factorization field, $B_i$,
couples to all fermion pairs, including to-be-free fermion pairs, $[\bar{\psi}_{R}(x)\psi_{iL}(y)]_f$:
\bbo
\label{r02}
S_{}(\Psi, B,Q)= W(\Omega)
+\!\int_x\bl  
 [\bar{\psi}_Li\slash{D}\psi_{L}]+ [\bar{\psi}_R\;i\slash{D}\psi_{R}]\br
 \nonumbo
 +\int_{xy}\; \bl -[B^i{}^\dagger D_F B_i]
+
\left( g_0 ( B^{i\dagger}+ Q^{i\dagger}) D_F[\bar{\psi}_{R}\psi_{iL}]_b + 
g_0 B^{i\dagger} D_F[\bar{\psi}_{R}\psi_{iL}]_f
+  h.c. \right)\br,
\ebo
where,
\bbo
\label{r03}
\int_x\bl  
 [\bar{\psi}_Li\slash{D}\psi_{L}]+ [\bar{\psi}_R\;i\slash{D}\psi_{R}]\br
 =\int_x\bl [\bar{\psi}_Li\slash{D}\psi_{L}]_b+ [\bar{\psi}_R\;i\slash{D}\psi_{R}]_b+[\bar{\psi}_Li\slash{D}\psi_{L}]_f+ [\bar{\psi}_R\;i\slash{D}\psi_{R}]_f\br.
\ebo
(Here we abbreviate integrals as:
\bbo 
\int_{u...v}\!\!\!\! =\int\!\!d^4u..d^4v\;,
\ebo
and we will suppress  writing field arguments where obvious, e.g.,
$B^i{}^\dagger(x,y) D_F(x,y) B_i(x,y)\rightarrow B^i{}^\dagger D_F B_i$,
etc.).

We then have: 
\bbo
\label{200}
\backo\backo
\frac{\delta S_{}(\Psi, B,Q)}{\delta Q^{i\dagger}} = g_0 D_F[\bar{\psi}_{R}\psi_{iL}]_b \equiv g_0\sqrt{N_c}M_0^2 D_F H_i ;\qquad   
\frac{\delta S_{}(\Psi, B,Q)}{\delta Q_i} = g_0  D_F[\bar{\psi}^i_{L}\psi_{R}]_b\equiv g_0\sqrt{N_c}M_0^2D_F H^{i\dagger};
\ebo
where we anticipate the  normalization of $H_i$,
$[\bar{\psi}_R(x)\psi_{iL}(y)]
\rightarrow [ \bar{\psi}_R(x)\psi_{iL}(y)]_{f}
+ M_0^2{\sqrt{N_c}} H_i(x,y) $, from ref. \cite{main}.\footnote{Only the 
color singlets bind, hence with color indices $(a,b)$,
 \bbo
\label{act2}
\backo\!\!\!\!\!\!\! \bar{\psi}^{a}_{R}(x)\psi_{ibL}(y)
\rightarrow  \bar{\psi}^{a}_R(x)\psi_{ibL}(y)_{f}
+ M_0^2\frac{\delta{^a_b}}{\sqrt{N_c}} H_i(x,y).
 \ebo
 which yields the $\sqrt{N_c}$ factors in eq.(\ref{200}).
 Note that in \cite{main} we used $M^2=\epsilon M_0^2$ as an arbitrary normalizing mass scale
 for $H$, but we then 
 showed that $\epsilon\rightarrow 1$ in the critical coupling limit.  We are
 presently abbreviating the discussion and assume $\epsilon=1$ at the outset.}

We perform the  Legendre transformation, where $Q$ is now viewed as a  functional of $H$
by eq.(\ref{200}), following \cite{CJT}:
\bbo
S(\Psi, B, H)= S_{}(\Psi, B,Q)- \int_{xy} \bl Q^\dagger  \frac{\delta S}{\delta Q^\dagger} + Q \frac{\delta S}{\delta Q}
\br,
\ebo
hence,
\bbo
\backo\backo
\label{24}
S(\psi, B, H)= W(\Omega)
+\!\int_x\bl  
 [\bar{\psi}_Li\slash{D}\psi_{L}]+ [\bar{\psi}_R\;i\slash{D}\psi_{R}]\br
 +\int_{xy}\; \bl -[B^i{}^\dagger D_F B_i]
+g_0(B^{i\dagger}D_FC_i +  h.c.) \br,
\ebo
where we define,
\bbo
\label{C}
C_i(x,y) = \sqrt{N_c}M_0^2 H_i(x,y)+ [\bar{\psi}_{R}(y)\psi_{iL}(x)]_f.
\ebo
If we ignore the quantum corrections ($\omega$-loops), 
the ``equation of motion'' of $B_i$ becomes:
\bbo
\label{qc3}
\backo\backo
B_i(x,y)=g_0C_i(x,y),\qquad  B^{i\dagger}(x,y)=g_0C^{i\dagger }(x,y).
\ebo
Substituting  $B_i$  into eq.(\ref{24})
we thus obtain the semiclassical result,
\bbo
\label{resultcccc}
S(\Psi, H)=
\!\int_x\! \bl  
 [\bar{\psi}_L(x)i\slash{D}\psi_{L}(x)]_b+ [\bar{\psi}_R(x)\;i\slash{D}\psi_{R}(x)]_b+
 [\bar{\psi}_L(x)i\slash{D}\psi_{L}(x)]_f+ [\bar{\psi}_R(x)\;i\slash{D}\psi_{R}(x)]_f
 \br
 \nonumbo
 +g_0^2\!\int_{xy} \bl
  N_cM_0^4 H^\dagger(x,y) D_F(x-y)H(x,y)  
 +\;\sqrt{N_c}M_0^2(H^{i\dagger}(x,y)D_F(x-y)[\bar{\psi}_{R}(y)\psi_{iL}(x)]_f+h.c.)
 \nonumbo \qquad\qquad
 + [\bar{\psi}^i_L(x)\psi_{R}(y)]_fD_F(x-y)[\bar{\psi}_{R}(y)\psi_{iL}(x)]_f\br.
\ebo
We have therefore obtained  results identical to the starting point of our
previous theory, eq.(54), in ref.\cite{main}. Note the BCS-like enhancement factor of $N_c$
in the $H^\dagger D_F H$ interaction term, where $N_c$ is the analogue of the 
number of Cooper pairs in a BCS state \cite{BCS}.

\subsection{Kinetic Terms and Factorization Ansatz}

The ``to-be-bound'' fermion kinetic terms remain in the action $S'$:
\bbo
\label{kt}
S_{KTb}=
\!\int_x\! \bl  
 [\bar{\psi}^i_L(x)i\slash{D}\psi_{iL}(x)]_b+ [\bar{\psi}_R(x)\;i\slash{D}\psi_{R}(x)]_b\br,
\ebo
These have to be replaced by the kinetic term of the bound state field $H_i(x,y)$.
 From eq.(\ref{200}), in the free field limit (small $g_0$), the kinetic terms $S_{KTb}$ 
 imply the equations of motion. Technically with $H\equiv H_i\frac{\tau^i}{2}$ these are best written as commutators, 
\bbo
[D_y,[D_y,H(x,y)]]=0\qquad [D^\dagger_x,[D^\dagger_x,H(x,y)]]=0.
\ebo
 Omitting complications from the gauge fields these follow from the square of the Dirac
equation, e.g. $[\bar{\psi}_{L}(y)[\slash{\partial}, [\slash{\partial},\psi_{iR}(x)]]_b
= \partial_x^2 [\bar{\psi}_{L}(y)\psi_{iR}(x)]]_b=0$, etc.
A free field Lorentz invariant action yields the equations of motion by variation:
\bbo 
\label{first}
\backo
S_{KTb}\rightarrow S_{KTH} =
M_0^4\int\! d^4x \; d^4y \; \bl Z|D_R^\dagger H(x,y) |^2
+ Z|D_L H(x,y)|^2 \br,
\ebo
where the covariant derivatives are as defined in the NJL model:
\bbo
 \backo\backo
D_{L\mu}=\frac{\partial}{\partial y_\mu}- ig_2W_\mu ^A(y)\frac{\tau^A}{2}-ig_1 W_\mu^0(y) \frac{Y_L}{2};
 \qquad 
 D^\dagger_{R\mu}=\frac{\partial}{\partial x_\mu}+ig_1 W_\mu^0(x) \frac{Y_R}{2},
\ebo 
where $W^A_\mu$ ($W^0_\mu$) denotes the $SU(2)$ {$U(1)_Y$ gauge fields and the weak hypercharges are
$Y_L= 1/3$, $Y_{tR}=4/3$, (and the presently unused $Y_{bR}=-2/3$),
and we will require $Y_H= -1$ for the BEH boson of the SM;
the electric charges are as usual: $Q=I_3 + \frac{Y}{2}$.
Note that $D_L$ ($D^\dagger_R$) acts at coordinate  $y$ ($x$), and $D^\dagger_R$ acts on $\bar{t}_R$,
hence the sign flip in the gauge field terms (the derivative $D^\dagger$ acts in the forward
direction as we have written the kinetic term eq.\ref{first}).

We now pass to barycentric coordinates,\footnote{ Note: the Jacobian was misquoted in 
the preprint version of \cite{main} but corrected in the published version. }
\bbo
\label{bary}
\backo\backo
X^\mu=\frac{x^\mu+y^\mu}{2},\;\;\; r^\mu=\frac{x^\mu-y^\mu}{2}, 
\;\;\;
\partial_x=\half(\partial_X+\partial_r),
\;\;\;
\partial_y=\half(\partial_X-\partial_r)
\;\;\;\makebox{Jacobian:}\;\;\; \left|\frac{\partial(X,\rho)}{\partial(x,y)}\right|\equiv J^{-1} = 2^{-4}.
\ebo

To work in the rest frame as in 
 \cite{main} we follow Yukawa, \cite{Yukawa}, and for
 the ground state we  consider a factorized ansatz for  $H_i(X,r)$:\footnote{Note that we {\em have not} specified that $H(x,y)$
 or $Q(x,y)$ is factorized in $(x,y)$, as $\sim \bar{\psi}_R(x)\psi_L(y)$, hence we are permitted
 to make the ansatz $H_i(X)\phi(r)$.  An $(X,r)$ factorized form is a mixture of $(x,y)$ factorized basis
 functions, e.g.,  $H(X)\phi(r)\sim \sum_{ij} a_{ij}\bar{\psi}_{Ri}(x)\psi_{Lj}(y)$, i.e., the boundstate is an entangled
 combination of free scattering states.}
 \bbo
\label{af1000}
H_i(x,y) = H_i(X+r,X-r)\equiv \hat{H}_i(X,r) \qquad \text{hence,} \qquad 
\sqrt{J/2}\;\hat{H}_i(X,r)= H_i(X)\phi(r).
 \ebo 

We can introduce  Wilson lines {\em as an approximation}  to ``pull-back''
the gauge couplings from $(x,y)$ to the center $X$. This is
done by field redefinitions (as discussed in \cite{main}):
\bbo
H_i(x,y)\rightarrow W^\dagger_R(X,x)W_L(X,y)\;\sqrt{2/J}\; H_i(X)\phi(r),
\ebo
where $H_i(X)$ will have mass-dimension 1, and and $\phi$ will be dimensionless.
Note that $H_i(X)$, is distinguished from $H_i(x,y)$ by having dependence only upon $X$,  
and will become the effective BEH boson.
The Wilson lines are not necessary.
In the pure gauge limit these are simply gauge transformations on the $H_i$ field.
We  
introduce the Wilson lines as a conceptual tool to
consolidate the electroweak charges at $X$, though the gauge charges are effectively
consolidated at $X$  in the low energy approximation.
Moreover, 
the Wilson lines are renormalizable and act as
sources (charges)  only at their end-points. However, 
they do radiate, acting as currents.  

With the pull-back localization of the BEH  electroweak charges, the $W$ and $Z$ masses 
are generated in the usual way under spontaneous symmetry breaking.\footnote{We have 
neglected the gluon Wilson line that extends from $-\vec{r}$ to $\vec{r}$, but
of course, produces no color charge of the BEH boson.}
We have the Lorentz invariant kinetic term action:
\bbo
\label{Haction}
 S_{KTH}\rightarrow \!\!\int\!\!d^4Xd^4r\; Z M_0^4 \bl |D_{H}H(X) |^2|\phi(r)|^2
+|H(X)|^2  |\partial_{r^\mu}\phi(r) |^2  \br,
 \ebo      
where  the covariant derivative is:
 \bbo
 \label{covD}
 D_{H\mu}=\frac{\partial}{\partial X^\mu}- ig_2W^A_\mu(X)\frac{\tau^A}{2}-ig_1 W^0_\mu(X) \frac{Y_H}{2}.
 \ebo
 The internal wave-function $\phi(r)$
is now a complex scalar that carries no gauge charges, and $Y_H=-Y_{tR}+Y_{L}=-1$, apropos the BEH field.
(see Appendix A of \cite{main} for the algebra of the Wilson line pullback).

\subsection{Removing Relative Time}

Consider the kinematics of a pair of massless particles of 4-momenta $p_1$ and $p_2$.  We have $p_1^2=p_2^2=0$.
and two-body plane waves,
$H(x,y) \sim \exp(ip_1x+ip_2y)$. We  pass to the total
momentum $P=(p_1+p_2)$ and relative momentum  $Q=(p_1-p_2)$,
and the plane waves become $\exp(iPX+iQr)$.
Note that $P_\mu Q^\mu=p_1^2-p_2^2=0$. 
This implies in the rest (barycentric) frame,
$P=(P_0,0)$ and $Q=(0,\vec{Q})$, and the free equation of motion, $P_0^2-\vec{Q}^{\;2}=0$. 
Hence, in the rest frame the dependence upon $\vec{X}$ (associated with $\vec{P}$) and, in particular 
the relative time, $r^0$ (associated with $Q^0$), drops out.
If the particles are constituents of a bound state then this is the rest
frame of the composite particle.   In any other frame there is always a boosted $r^0{}'$
that is unphysical.

Bilocal fields have spurious dependence upon $r^0$ in the rest frame
(or its boosted $r^0{}'$ in any other frame).  This is an unphysical and unwanted
degree of freedom. It is analogous to the situation in a gauge theory where $\partial_\mu A^\mu$
is unphysical.\footnote{In a gauge theory, such as QED, one may think this is relevant
only to define the path integral. However this is an issue in classical electrodynamics, as
well, and leads to the ``transverse current'' in general gauges \cite{Jackson} }  
To use the bilocal formalism, $r^0{}'$
must be projected out with Lorentz invariant constraints, analogous to gauge fixing. 
We have discussed this in greater
detail in the present action formalism elsewhere \cite{main,main2}.

In the action of eq.(\ref{Haction})
we included a normalization factor $Z$ as in \cite{main}.  This
is natural in mapping from the fermionic kinetic terms to the $\hat{H}_i$ 
kinetic terms, but it serves the role of allowing us
to remove relative time.  
In the barycentric (center-of-mass) frame we assume an ansatz for
$\phi(r)$ ($H_i(X))$ that has no dependence upon the relative
time $r^0$ ($\vec{X}$). The field $H_i(X)$ will have the conventional volume normalization
of a mass-dimension 1, complex field.
%
The field $\phi(\vvr)$ however,  will describe the bound state
and must be a compact (normalizable) field.  

A canonical normalization of the $H_i(X)$ kinetic term in any frame (which leads to
a normalized Noether current, such as 
 $iH^\dagger \overleftrightarrow{\partial}_{\!\!\!X} H$)
dictates a Lorentz invariant normalization constraint:
  \vspace{-0.0in}
\bbo \label{basenorm}
 1= \int  d^4r\; ZM_0^4\; |\phi(r)|^2 .
\ebo
Since  the relative time
disappears kinematically in the rest frame, then  $\phi(r)\rightarrow \phi(\vvr)$ becomes a static field that
has no dependence upon $r^0$. 
We then  define the compact normalization for $\phi(\vvr)$ in the rest frame as:
  \bbo 
 \label{phinorm}
  1=  M_0^3 \int d^3 r\; |\phi(\vec{r})|^2,  
  \qquad  \text{which implies,} \qquad
\int ZM_0\; dr^0=1, \qquad \text{(see below).}
\ebo
These conditions can be specified in a manifestly covariant form using
the unit vector $\omega_\mu = P_\mu/\sqrt{P^2}$, 
and e.g., $ r^0 = r_\mu \omega^\mu $,  etc.,  as in Appendix C of \cite{main}.
Note that $\phi(\vvr)$ is  dimensionless with eq.(\ref{phinorm}).

This can be viewed as a ``prenormalization''  to define
the free bilocal field and it's Noether charges \cite{main2}, before turning on the interaction.
The factor $Z$ {\em is not absorbed into the fields} after turning on the interaction
since we don't want the
coupling constant, $g_0^2$, to be renormalized at the classical level. Indeed, 
$Z$ can be viewed as an operator, $Z\sim \delta(M_0r^0)$.
We remark that all field theories, even local ones, are implicitly defined 
with a  ``canonical'' prenormalization that defines the semiclassical action
after which we turn on the interaction. Then, at loop level, we are compelled to
renormalize the fields to preserve the canonicality which, in turn, renormalizes the coupling constants.

We see that $S_{KTH}$ becomes,
\bbo
\label{SKTH}
S_{KTH} \rightarrow \!\!\int\!\! d^4X \bl
 |D_XH(X)|^2 
+  
|H(X)|^2 M_0^3\!\!\int \!\! d^3r\; \left(
-|\partial_{\vvr}\phi(\vvr)|^2 \right)\br.
\ebo
We then turn on the interaction with coupling constant 
$g_0$ of eq.(\ref{resultcccc}), and
the semiclassical action becomes,
  \bbo
  \label{intD}
\backo\backo \!\! S_0(H) =
\!\!\int\!\! d^4X \bl
 |D_XH(X)|^2 
+  
|H(X)|^2 M_0^3\!\!\int \!\! d^3r\; \left(
-|\partial_{\vvr}\phi(\vvr)|^2 \right)
+|H(X)|^2\!\!\int\!\! dr^0 d^3r \left( M_0^4 \;{2g_0^2 N_c }D_F(2r^\mu)|\phi(\vvr)\; |^2\right)
\!\br.
 \ebo 
Note that there is no $Z$ factor in the interaction term, and
$|\partial_{r^\mu}\phi|^2= |\partial_{r^0}\phi|^2 - |\partial_{\vec{r}}\phi|^2\rightarrow - |\partial_{\vec{r}}\phi|^2$.
If we were to define $z\equiv \int ZMdr^0 $ , then we can rescale $\phi\rightarrow \phi/\sqrt{z}$ to
have the compact normalization of eq.(\ref{phinorm}),
but then the coupling constant becomes $g_0^2/z$.  However, we want the theory to have at the classical
level the same coupling $g_0^2$ as defined by the coloron exchange, e.g., we want to match the low energy scattering matrix elements
that we would obtain in the conventional topcolor field theory to the bilocal theory.  Hence $z=1$ is a matching condition of the 
underlying topcolor interaction to the bilocal theory. 

We therefore integrate over $r^0$ in the interaction term :
       \bbo\label{Yp}
\backo\!\!\!\int\!\! dr^0  D_F(2r)
= -\!\!\int\!\!  dr^0 \frac{d^4q}{(2\pi)^4}
\frac{1}{q^2-M_0^2}e^{2i q_\mu r^\mu }
= \half \int\!\!   \frac{d^3q}{(2\pi)^3}
\frac{1}{\vec{q}{\;}^2+M_0^2}e^{2i q_\mu r^\mu }= -\half V_0(2|\vvr|)=\frac{ e^{-2M_0 |\vvr|}}{16\pi |\vvr|}.
\ebo 
The $\vec{q}$ momentum integral yields the familiar Yukawa potential
(where $2|\vvr| $ is the separation of the particles).    
The action then becomes,
  \bbo
  \label{intDf}
\backo S =
\!\!\int d^4X \bl
 |D_XH(X)|^2 
+  
|H(X)|^2 M_0^3\!\! \int \!\! d^3r \left(-
|\partial_{\vvr}\phi(\vvr)|^2 
+{g_0^2 N_c  M_0}\frac{ e^{-2M_0 |\vvr|}}{8\pi |\vvr|}|\phi(\vvr) |^2
\right) \br.
 \ebo 
 While not ``manifestly' so, this is ``implicitly'' Lorentz invariant as discussed in Appendix C of \cite{main}.
 Note that, in the limit of suppressing the
$\vec{q}{\;}^2$ in the denominators of the integrands of eq.(\ref{Yp}), we obtain the large $M_0^2$ limit
of the potential
(using $J=2^4$, 
and $\delta^3(\vec{r})=(4\pi r^2)^{-1}\delta(r)$):
\bbo
\label{Ypot2}
V_0(2r)\rightarrow -\frac{1}{M^2_0}\delta^3(2\vvr)=-\frac{1}{8M^2_0}\delta^3(\vvr)
\qquad D_F(2r)\rightarrow \frac{1}{M^2_0}\delta^4(2r^\mu)=\frac{1}{J M^2_0}\delta^4(r^\mu).
 \ebo     
 This recovers the NJL model potential.
 
 The manipulation leading to eq.(\ref{intDf}) may seem to be inconsistent, but it is a short-cut for a more
 formal limiting procedure. In a manifestly
 Lorentz invariant form of the internal wave-function action of eq.(\ref{intD}),
 \bbo
  \label{intD1}
\!\!\int \!\! d^4r\; \bl
Z|\partial_{r^\mu}\phi(r^\mu)|^2 
+ \;{2g_0^2 N_c }D_F(2r^\mu)|\phi(r^\mu)\; |^2
\!\br,
 \ebo
 we want to take $\phi(r^\mu)\rightarrow \phi(\vvr)$,  which is independent of $r^0$. However
 this appears to be inconsistent with 
 the interaction term with the factor $D(2r^\mu)$ which has nontrivial dependence upon $r^0$.
 The problem arises from having integrated out non-static colorons
 to produce $D_F(2r^\mu)$. However to obtain a Hamiltonian (such as in functional Schr\"odinger picture)
 we go to static field configurations.  
 To see this, consider a toy constituent, complex scalar, field theory:
  \bbo
  \label{intD2}
 S=\!\!\int \!\! d^4x \bl(
|\partial\varphi_1(x)|^2 +|\partial\varphi_2(x)|^2  -gM_0(\varphi_1(x)\varphi_2(x)A(r)+h.c.)+|\partial A(x)|^2
-M_0^2 |A(x)|^.
\br.
 \ebo
The Hamiltonian for this theory takes the form
\bbo
H= \!\!\int \!\! d^3x \; \bl
|\pi_1(\vec{x})|^2+|\pi_2(\vec{x})|^2+ |\nabla\varphi_1(\vec{x})|^2 + |\nabla\varphi_2(\vec{x})|^2 
+gM_0(\varphi_1(\vec{x})\varphi_2(\vec{x})A(\vec{x})+h.c.)
\nonumbo \qquad\qquad
+|\pi_A(\vec{x})|^2 +|\nabla A(\vec{x})|^2+M_0^2 |A(\vec{x})|^2
\br,
\ebo
where the canonical momenta are $\pi_i=\partial_0\varphi_i$ and $\pi_A=\partial_0 A$.
We are interested in the static limit of this theory so we impose the
constraints,  $\pi_i=\pi_A=0$, and there
is then no Poisson bracket (hence no Dirac bracket). 
The static equation for $A(\vec{x})$ is, $\nabla^2A -M_0^2A =gM_0\varphi_1\varphi_2$,
hence we can integrate out $A$, and write,
\bbo
H= \!\!\int \!\! d^3x d^3y\; \bl
ZM_0|\nabla_x\varphi_1(x)|^2|\varphi_2(y)|^2+ZM_0|\nabla_y\varphi_2(y)|^2|\varphi_1(x)|^2 - g^2M^2_0 |\varphi_1(x)\phi_2(y)|^2\Delta(x-y) 
\br,
\ebo
where $1=ZM_0\int d^3y|\varphi_i(y)|^2$, and $\Delta(x-y)=V_0(\vvx-\vvy)/2$.
Now, replace $\varphi_1(x)\varphi_2(y)\rightarrow M_0\Phi(x,y)$  to   obtain,
\bbo
H= \!\!\int \!\! M_0^3 d^3x d^3y\; \bl
Z|\nabla_x\Phi(x,y)|^2+  Z|\nabla_y\Phi(x,y)|^2 - \half g^2M_0 |\Phi(x,y)|^2V_0(x-y) 
\br.
\ebo
We go to the barycentric coordinates, and factorize $\Phi(x,y)=\sqrt{2/J}\; \chi(X)\phi(\vvr)$,\footnote{A c-number
product $\varphi_1(\vec{x})\varphi_2(\vec{y})$ cannot generally be rewritten in the factorized form, but 
quantum fields can lead to states that are so factorized, $\langle 0|\varphi_1(\vec{x})\varphi_2(\vec{y})|S \rangle 
\sim  \chi(X)\phi(\vvr) $.} 
\bbo
H= \!\!\int \!\! M_0^3 d^3X d^3r\; \bl
Z|\nabla_X\chi(\vec{X})|^2|\phi(\vvr)|^2+ Z|\chi(\vec{X})|^2 |\nabla_{\vvr} \phi(\vvr)|^2 - g_0^2 N_c M_0 |\chi(\vec{X})|^2 |\phi(\vvr)|^2V_0(2\vvr)
\br ,
\ebo
 with $g^2 =g_0^2 N_c $.
 We can do the same construction with the full
coloron theory. The above example displays the origin of $Z$ and shows that there is no
inconsistency in the final result of eq.(\ref{intDf}) obtained by the short-cut 
which yields the single
particle exchange interaction as it would be generated in the  static frame.
This is the correct procedure leading to a Hamiltonian for the functional Schr\"odinger equation
of the field theory with the bound state (see the discussion in \cite{CJT}
of Kuti's method).  To recover the action formalism we would do a Legendre
transformation from $H$ back to the Lagrangian $L$,
restoring the time derivative
on $\chi(X)$ and integrating $L$ over $X^0$.

We emphasize that we can often use the NJL limit of the {\em potential}, but the bilocal field
remains.  For example, in the Dirac $\delta$-function potential of $1+1$ quantum
mechanics, the wave-function $\phi(r)$ is extended in space even though the potential
is pointlike \cite{main}.  In  $1+3$ dimensions 
the Dirac $\delta$-function potential theory is 
actually ambiguous, and 
depends upon the limiting procedure
used to define it.  This is the analogue of what we are facing
presently, but the Yukawa potential affords a well defined limiting
procedure and unambiguous solutions.

Including the Yukawa interaction 
and a loop generated quartic term  we obtain
the effective action for the composite BEH field,
  \bbo
\backo\backo S=
\!\!\int\!\! d^4X\;\bl
 |D_H H(X)|^2+ 
|H(X)|^2 \; M_0^3\!\! \int\!\! d^3r\; \bl
-|\partial_{\vec{r}} \phi(r)|^2 
+{g_0^2 N_c M_0}\frac{ e^{-2M_0 |\vvr|}}{8\pi |\vvr|}|\phi(\vvr) |^2\br 
\nonumbo\qquad\qquad
-\frac{\lambda}{2}(H^\dagger(X) H(X))^2 - g_Y\left([\bar{\psi}^i_{L}(X)t_{R}(X)]_{f}H_i(X)  +h.c.\right)\br.
   \ebo   
The internal field $\phi(r)$ is ``nested'' within the
action for a conventional pointlike BEH boson, $H(X)$. The static $\phi$ field
has a Hamiltonian:
\bbo
{\cal{M}}
=
M_0^3\!\! \int\!\! d^3r\; \bl
|\partial_{\vec{r}} \phi(r)|^2 
-{g_0^2 N_c M_0}\frac{ e^{-2M_0 |\vvr|}}{8\pi |\vvr|}|\phi(\vvr) |^2\br .
\ebo
Extremalization of the Hamiltonian ${\cal{M}} $ yields the Schr\"odinger-Klein-Gordon (SKG) equation
for $\phi(\vvr)$ with the eigenvalue $\mu^2$:
 \bbo
 -\nabla^2 \phi -g_0^2N_c M_0\frac{ e^{-2M_0 |\vvr|}}{8\pi |\vvr|}\phi(r) =\mu^2\phi.
 \qquad\text{where,}\qquad \nabla^2 =\frac{\partial^2 }{\partial r^2}+\frac{2}{r}\frac{\partial }{\partial r}.
 \ebo
We find that the SKG equation has
a critical coupling, $g^2_cN_c/8\pi^2 = 1.06940$, for which $\mu^2=0$. Remarkably,  this is very close to the quantum
NJL critical coupling  $g^2_{NJLc}N_c/8\pi^2 = 1.00$ (see Section IV. (B) of \cite{main}).  When $g^2_0>g^2_c$ the eigenvalue  $\mu^2$
becomes negative, $\rightarrow -|\mu|^2$.  In such a solution the action
for $H(X)$ then becomes the familiar,
\bbo
\backo\!\!\! S=
\!\!\int d^4X \bl
 |D_H H(X)|^2+|\mu|^2 |H(X)|^2
-\frac{\lambda}{2}(H^\dagger(X) H(X)) - g_Y\left([\bar{\psi}_L(X)t_{R}(X)]_{f}H(X)  +h.c.\right)\br,
   \ebo 
 with the ``sombrero potential'':
 \bbo
-|\mu|^2 (H^\dagger(X) H(X))+\frac{\lambda}{2}(H^\dagger(X) H(X))^2 .
   \ebo 

   As discussed in detail in \cite{main}, 
  the solution of the Schr\"odinger-Klein-Gordon SKG equation for $\phi(r)$ indeed extends to large distances, 
  $\phi(r) \sim e^{-|\mu|r}/r$ where $|\mu|<\!\!M_0$ near critical coupling.
  This dilutes the value of $\phi(0) \sim \sqrt{|\mu|/M_0}$. 
  We find that the Yukawa coupling $g_Y \propto \phi(0)$
  and $\lambda \propto g_Y^4 \propto  |\phi(0)|^4$.
  Moreover, inputting the known value of the Lagrangian mass of the BEH boson
  in the symmetric phase, 
  $-|\mu|^2 =-(88)^2$ GeV$^2$, we find the composite scale (coloron mass)
  $M_0\approx 6$ TeV.
  The quartic coupling, $\lambda$, will be determined at loop level
  and we find remarkable agreement with the SM as discussed below.
  
  The degree of fine-tuning of the theory is also suppressed by $\phi(0)$ in
  a subtle way. Rather than the naive result one would expect from the NJL model, $\delta g_0^2/g_c^2 \sim|\mu|^2/M_0^2 \sim 10^{-4}$,
  we now obtain a linear relation: $\delta g_0^2/g_c^2 \sim|\mu|/M_0 \sim 1\%$.  To obtain these results we need
  the precise relationship between $g_Y$ and $\phi(0)$ which follows from the induced 
  Yukawa coupling of the bound state to free fermions.

 \subsection{The Induced Bound State Yukawa Interaction}
 
We see in eq.(\ref{resultcccc}) that the Yukawa interaction of the bound state with
the free scattering state fermions is now induced from the $\sqrt{N_c}$ term:
\bbo 
\label{5NJL000}
\backo
 g^2_0  \sqrt{2N_cJ} M_0^2
\int\!\! d^4X d^4r \;[\bar{\psi}^i_{L}(X\!+\!r)\psi_{R}(X\!-\!r)]_{f}D_F(2r)\;H_i(X)\phi(\vvr) {+h.c.}.
\ebo
Consider the pointlike limit of the potential of eq.(\ref{Ypot2}),
 $ D_F(2r)\rightarrow (JM_0^2)^{-1} \delta^4(r)$:
\bbo
\rightarrow
g^2_0  \sqrt{2N_cJ}
\int\!\! d^4X\; [\bar{\psi}^i_L(X)\psi_{R}(X)]_{f}H_i(X)\phi(0) {+h.c.}.
\ebo
We therefore see that the induced Yukawa coupling to the field $H(x)$ is:
\bbo
\label{gy}
g_Y= g_0^2\sqrt{2N_c/J}\;\phi(0).
\ebo
 Here the behavior
of $\phi(0)$ is a suppression of $g_Y$ that is
a
power-law, $\sim \sqrt{|\mu|/M_0}$ for small $\mu$,  near the critical coupling.
This power-law behavior is significantly different than the slow
RG (logarithmc) evolution in the old NJL-based top condensation model, which is why we now obtain $M_0\sim 6$ TeV
and reduced fine-tuning $\sim |\phi(0)|^2\sim$ few $\%$ (rather than $M_0\sim 10^{15}$ GeV and extreme fine-tuning
$\sim 10^{-26}$).

\section{Quantum Fermion Loop Corrections to the Semiclassical Theory \label{quantum}}

We recall the results for the fermion loops that arise in the pointlike NJL model with Yukawa interaction:
\bbo
\label{yuknjl}
\int_x \bl g_0[\bar{\psi}^i_{L}(x)\psi_{R}(x)]H_i(x)+h.c. \br.
\ebo
 This leads to the loop-induced,  ${\cal{O}}(\hbar)$, terms in an effective action (ignoring classical terms):
\bbo
\label{0NJL3}
\backo
S_{eff}
=\!\int\! d^4x \bl \widetilde{Z} DH^\dagger DH + M^2H^\dagger H - \frac{\lambda }{2}(H^\dagger H)^2  
\br,
\ebo
where,
\bbo
\label{0NJL4}
\backo
\widetilde{Z}=\frac{g_0^{2}N_{c}}{8\pi ^{2}}\ln(M_0/m), \qquad
M^2 = \!\frac{g_0^{2}N_{c}}{8\pi ^{2}}M_0^{2},
\qquad
\lambda=\frac{g_0^{4}N_{c}}{4\pi ^{2}}\ln( M_0/m).
 \ebo
The sign of the loop-induced $M^2$ term in the action is positive, representing an attractive negative term
(tachyonic mass)
in the effective potential, while the quartic term is negative in the action, hence, repulsive
in the potential.  In the NJL model, with classical bare mass $M_0^2$, the resulting physical
mass of the pointlike bound state is $\mu^2 = \widetilde{Z}^{-1}( M_0^2 - M^2) $,
which defines the NJL critical coupling, $1=g_0^2 N_c/8\pi^2$. Hence $M^2$ is
the analogue of ``binding energy'' associated with the bound state.

This is a hint that the quantum loop effects will enhance the binding of
the interaction in the semiclassical bilocal theory.  
However, it is important to allow the internal wave-function to respond to the 
attractive $M^2$ term, which now becomes part of the potential.
Moreover,  the NJL limit is  a $\delta$-function approximation, and we 
expect in the bilocal theory a softer contribution to the short distance potential than 
the mass $M^2$. 
Indeed, in the the loop calculation the term  $M_0^2\delta^4(2r)$  is indistinguishable from
the softer $D_F(2r)$ in the bilocal theory.  Hence the interpretation of the NJL $M^2$ as a 
loop correction to the mass
is now replaced in the bilocal theory as a loop correction to the potential, enhancing binding.

Hence, in the present analysis we will use the NJL limit in obtaining approximate bilocal loop integral results,
but judiciously replacing 
$\delta^4(x-y)\rightarrow M_0^2D_F(x-y)$. Physically,  the bilocal  theory has a true
momentum space cutoff of loop integrals $\sim M_0^2$,
due to $D_F(q^2)$, while in the NJL theory this cut-off is implemented in loop integrals ``by hand.''
The replacement of the $\delta^4(x-y)$  by  $M_0^2D_F(x-y)$ approximately implements the true physics of
the bilocal theory, allowing us to obtain a potential that, in turn, allows the ``back-reaction'' of the solution.
We are invoking the ``indistinguishability'' of the  $\delta^4(x-y)$  from the bilocal $M_0^2D_F(x-y)$,
but maintaining the bilocality of  the present theory.
By extending the auxiliary field equation to include the additional terms of $ W(\Omega) $,
then  integrating out $B_i$,
we obtain the quantum loop corrections to the semiclassical action.  These are the
analogues of the terms in eqs.(\ref{0NJL3},\ref{0NJL4}) generated by eq.(\ref{yuknjl}), 
but now for the bilocal fields.

To compute $W(\Omega)$ we compute the path integral of eq.(\ref{133}), integrating out $\Omega$. 
The coupling of $B_i$ to the underlying $\omega$-fields is implicit in
eqs.(
\ref{7},\ref{qbb}):
\bbo
\int_{xy}\bl
g_0\bar\omega_R(y)D_F(x-y)\omega_{iL}(x)B^{i\dagger}(x,y)+ h.c.\br.
\ebo
This yields single-particle irreducible 
(1PI) diagrams leading to quadratic and quartic terms in $B$:  
\bbo
\label{newaction2}
\backo\backo
\frac{i}{\hbar}W(\Omega)=
\!\int_{xyx'y'}\!\!\!\! 
[B^\dagger(x,y){\cal{F}}(x,y,x',y') B(x',y')],
\nonumbo 
+ \!\int_{x...z'}\!\!\!\! 
[B^\dagger(x,y)B(x',y')]   {\cal{G}}(x,y,x',y',w,z,w',z')[B^\dagger(w,z)B(w',z')] +...
\ebo
which largely parallels the loop corrections of the NJL model. 
We give the final results of the loop
calculations below in eqs.(\ref{resultaaa},\ref{result0}).

\subsection{Main Loop}

Note the first   term of eq.(\ref{newaction2}) contains  the 4-point function ${\cal{F}}(x,y,x',y')$,
given by:
\bbo
\backo\backo
\frac{i}{\hbar}W_{\cal{F}}(\Omega)=
\!\int_{xyx'y'}\!\!
[B^\dagger(x,y){\cal{F}}(x,y,x',y') B(x',y')]=
\nonumbo
-\half g_0^2 N_c\!\int_{x...y'}\!\!\![B^\dagger(x,y)
 D_F(x-y)\Tr(S_F(x-x')S_F(y'-y))D_F(x'-y')B(x',y')],
\ebo
where $S_F(x)$ is a fermionic Feynman propagator (we use conventions of \cite{BjDrell}):
\bbo
S_F(x) = \int \frac{d^4\ell}{(2\pi)^4}\frac{i\slash{\ell}}{\ell^2+i\epsilon}e^{i\ell\cdot x}
\qquad \makebox{and} 
\qquad \Tr\bl \frac{(1-\gamma^5)}{2}\slash{\ell}\;\slash{\ell}'\br = 2\ell\cdot\ell'.
\ebo
In the loop integral we make the NJL approximation
of taking a pointlike limit of the integral, defined by:
\bbo
D_F(x-y)=-\int \frac{d^4q}{(2\pi)^4}\frac{e^{iq(x-y)}}{q^2-M_0^2}\rightarrow
\int \frac{d^4q}{(2\pi)^4}\frac{e^{iq(x-y)}}{M_0^2}=\frac{1}{M_0^2}\delta^4(x-y),
\ebo
then,
\bbo
\backo\backo
{\cal{F}}(x,y,x',y') \rightarrow
%
-2g_0^2 N_c \frac{\delta^4(x-y)\delta^4(x'-y')}{M_0^4} 
\int\frac{d^4\ell}{(2\pi)^4}\frac{d^4\ell'}{(2\pi)^4}
\frac{ \ell\cdot\ell' e^{i(\ell-\ell')(x-x')} }{\ell^2\ell'^2}.
\ebo
It is useful to go to barycentric coordinates:
 \bbo
\label{barycentric}
X=\half(x+y), \;\;\; r=\half(x-y),
\;\;\; J=\left|\frac{\partial(x,y)}{\partial(X,r)}\right|=2^{4},
\;\;\;
\ebo
and define:
\bbo
\widetilde{B}(X,r)\equiv B(X-r,X+r). 
 \ebo
 We will also define new integration variables,
$\bar{X}=(X+X')/2$ and $R=(X-X')$, which have unit Jacobian.
 
Hence we have, with $\delta^4(x-y)=J\delta^4(2r) $,
\bbo
\backo\backo
\!\int_{xyx'y'}\!
[B^\dagger(x,y){\cal{F}}(x,y,x',y') B(x',y')]
\nonumbo
\backo
\rightarrow
-g_0^2 N_c J^2\!\int_{XrX'r'}\!\!\!\!\! [\widetilde{B}^\dagger(X,r)
\frac{\delta^4(2r)\delta^4(2r')}{M_0^4} 
\int\frac{d^4\ell}{(2\pi)^4}\frac{d^4\ell'}{(2\pi)^4}
\frac{ 2\ell\cdot\ell' e^{2i(\ell-\ell')(X+r-X'-r')} }{\ell^2\ell'^2}\widetilde{B}(X',r')]
\nonumbo
\backo
=
-\frac{g_0^2 N_c}{M_0^4}\!\int_{XX'}\!
 [\widetilde{B}^\dagger(X,0)   \int\frac{d^4\ell}{(2\pi)^4}\frac{d^4\ell'}{(2\pi)^4}
\frac{ 2\ell\cdot\ell' }{\ell^2\ell'^2}e^{i(\ell-\ell')(X-X')}           
 \widetilde{B}(X',0)].
\ebo
Note the cancellation of $J^2=2^8$ with the $(1/2^4)^2$ coming from 
integrating  $\delta^4(2r)\delta^4(2r')$.
Then  $\widetilde{B}^\dagger(X,0)$  ($\widetilde{B}(X',0)$) is reduced to a pointlike field with dependence only
upon $X$ ($X'$) where $r,r'\rightarrow 0$.

An outgoing (incoming) state is, $\widetilde{B}^\dagger(X,0)\sim \widetilde{B}^\dagger_0e^{iP(\bar{X}+R/2)}$,
( $\widetilde{B}(X',0)\sim \widetilde{B}_0e^{-iP(\bar{X}-R/2)}$).
Thus, integrating over $(\bar{X},R)$ we obtain,
\bbo
\label{resultb}
-\frac{g_0^2 N_c}{M_0^4}
\!\int_{{X}R} e^{i(\ell-\ell'+P)(R)}
 [\widetilde{B}^\dagger_0{}'\widetilde{B}_0]   \int\frac{d^4\ell}{(2\pi)^4}
\frac{ 2\ell\cdot(\ell') }{\ell^2(\ell')^2} 
= -\frac{2g_0^2 N_c}{M_0^4}
\!\int_{{X}}
 [\widetilde{B}^\dagger_0{}'\widetilde{B}_0]   \int\frac{d^4\ell}{(2\pi)^4}
\frac{ \ell\cdot(\ell + P) }{\ell^2(\ell+P)^2}          
\nonumbo
\qquad\qquad
=\frac{ig_0^2 N_c}{8 \pi^2 M_0^4}\!\int_{\bar{X}}[\widetilde{B}_0^\dagger \widetilde{B}_0]\bl M_0^2 + P^2 \ln\left(\frac{M_0}{m}\right) \br.
\ebo
Here we use a Wick rotation and  Euclidean momentum UV cut-off, $\ell^2< M_0^2$.

We have obtained the analogues of the $M^2$ and $P^2$ terms in eqs.(\ref{0NJL3},\ref{0NJL4})
acting on the auxiliary field.
The $P^2$ term corresponds to a loop-generated kinetic term for
the auxiliary field $B_i$, (see eq.(\ref{res})).
In the framework of the NJL model and original top condensation theory,
\cite{BHL}, this corresponds the formation of a physical state, i.e., a resonance.
We are interested presently in small $P^2$ so we will treat the
$P^2\approx 0$ limit of the loop and return to this issue below.

The remaining  $M_0^2$ term
of eq.(\ref{resultb}) is just the quadratically 
divergent two-point loop of the NJL model in disguise, as in eq.(\ref{0NJL4}).
We multiply by $-i\hbar$ to obtain the two-point loop contribution to the  action $W_{\cal{F}}(\Omega)$,
\bbo 
W_{\cal{F}}(\Omega) =\frac{g_0^2 N_c}{8 \pi^2 M_0^2}\!\int d^4X  \; [\widetilde{B}^\dagger(X,0) \widetilde{B}(X,0)]
\rightarrow   \frac{g_0^2 N_c J}{8 \pi^2 M_0^2}\!\int d^4X d^4r \; \widetilde{B}^\dagger(X,r) \delta^4(2r) 
\widetilde{B}(X,r).
\ebo
Note the positive sign in the action, corresponding to an attractive potential.
While in the pointlike NJL model this term was interpreted as the quadratic negative
term in the potential, $-g^2 N_c M_0^2/8\pi^2$,
we see that, in the bilocal theory due to the  implicit $\delta^4(2r)$, {\em this is a potential and not a mass term.}
This presently  acts as an attractive $\delta^4(2r)$ potential,
and the wave-function $\phi(r)$ will react to it as such. However, as stated above,
the $\delta^4(2r)$ is indistinguishable from the tree-level potential via:
$
\delta^4(2r)\rightarrow M_0^2D_F(x-y).
$
Therefore, using this replacement of the pointlike $\delta$-function
we obtain in the action $(P^2\approx 0)$:
\bbo
W_{\cal{F}}(\Omega) =
\!\int_{x...y'}\!\!\!
B^\dagger(x,y){\cal{F}}(x,y,x',y') B(x',y')
\approx
\frac{g_0^2 N_c}{8 \pi^2 }\int d^4x \; d^4y \; B^\dagger(x,y)D_F(x-y) B(x,y),
\ebo
where we have returned to the original $(x,y)$ coordinates, and we have
a two-body interaction.  This is loop level, ${\cal{O}}\hbar$, and upon integrating out the auxiliary field, $B$, this
will yield an enhancement of the semiclassical potential. 

This result answers the potential criticism of \cite{main}, mentioned
in the Introduction: If one computes the 
loop correction to the BEH mass by naively looping the Yukawa interaction in the
semiclassical theory, then
one would obtain the result of eq.(\ref{resultb}), which 
appears to be a large correction to the BEH mass $\propto M_0^2$.
Of course, this would be falsely double counting, since
the only occurrence of the loop is in the underlying theory.
Moreover,  
 as emphasized above, {\em this is not a mass term for the bilocal field since
it is $\propto\ \delta^4(2r)$ and, rather, it is
a short-distance (attractive) correction to the potential for $\phi(r)$.}  The $\phi(r)$ field will
adjust accordingly as the solution to the SKG equation, multiplicatively modifying the eigenvalue $\mu^2$.
Indeed, due to effective large distance scale symmetry near
critical coupling, this must be interpreted as a radiative 
(multiplicative) correction to the effective coupling in the semiclassical potential. 
The loop integral leads to an attractive interaction which represents an enhancement of the effective
coupling.  We elaborate this further below.

\subsection{Quartic Coupling}

We also obtain the induced 
quartic interaction, which appears as an 8-point function
in the effective semiclassical theory.
This can be likewise  evaluated in the NJL approximation.
The loop integral involves $D_F^4(2r) \Tr{(S_F)^4} $ and
we replace three of the  $D_F$ factors with $\delta^4(2r)/M_0^2$
which leaves one two-point integral of order $g_0^4$.
The result can be inferred from the NJL calculation of the $\lambda$ term
in eq.(\ref{0NJL4}) and is consistent with the detailed calculation:
\bbo
\backo\backo
W_{\cal{G}}(\Omega) =
\int_{x...z'}\!\!\!\! \!\!\!\!
B^\dagger(x,y)B(x',y') {\cal{G}} (x,y,x',y',w,z,w',z')B^\dagger(w,z)B(w',z')
\nonumbo
\qquad 
\approx
-\frac{{g}^4_0\hat\lambda }{2M_0^8}
\bl\!\int_{xy}\!\! ({B}^\dagger(x,y){B}(x,y))^2 \delta^4(x-y)\br,
\ebo
where we define:
\bbo
\label{hatlam}
\;\; \hat\lambda= \frac{N_c}{4\pi^2}\ln\bl\frac{M_0}{m}\br.
\ebo
We develop this further below.

\subsection{Full Renormalized Action}


The resulting action with eq.(\ref{SKTH}) and the principal quantum effects 
considered here is then :
\bbo
\backo\backo
\label{240}
S= S_{KTH}+W(\Omega)
+\!\int_x\bl  
 [\bar{\psi}_Li\slash{D}\psi_{L}]_f+ [\bar{\psi}_R\;i\slash{D}\psi_{R}]_f\br
+\int_{xy}\; \bl -[B^i{}^\dagger D_F B_i]
+
g_0(B^{i\dagger}D_F C_i+h.c.) \br.
\ebo
We recall eq.(\ref{C}), $ C_i(x,y) = \sqrt{N_c}M_0^2 H_i(x,y)+ [\bar{\psi}_{R}(y)\psi_{iL}(x)]_f$,
and we have from the loop calculations:
\bbo
\label{newaction20}
\backo\backo
W(\Omega)=
\!\int_{xyx'y'}\!\!\!\! 
[B^\dagger(x,y){\cal{F}}(x,y,x',y') B(x',y')]
+ \!\int_{x...z'}\!\!\!\! 
[B^\dagger(x,y)B(x',y')]   {\cal{G}}(x,...,z')[B^\dagger(w,z)B(w',z')] +... ,
\nonumbo
=\frac{g_0^2 N_c}{8 \pi^2 }\int_{xy} \; B^\dagger(x,y)D_F(x-y) B(x,y)
-
\frac{g_0^4\hat\lambda}{2M_0^8}\!\int_{xy}\!\! ({B}^\dagger(x,y){B}(x,y))^2 \delta^4(x-y)  .
\ebo
Hence $S$ can be can be rewritten as: 
\bbo
\label{resultaaa}
\backo\backo
S
=S_{KTH}+\!\int_x\bl [\bar{\psi}_Ri\slash{D}\psi_R]_f+[\bar{\psi}_Li\slash{D}\psi_L]_f  \br
+ \!\int_{xy} \!\bl g_0 B^{i\dagger} (x,y)D_F(x-y)C_i+h.c.\br
\nonumbo
-\!\int_{xy} \!
[B^\dagger(x,y) D_F(x-y)B(x,y)]\bl 1 -\frac{g_0^2 N_c}{8 \pi^2 }\br
- \frac{g_0^4\hat\lambda}{2M_0^8}\int_{xy}\!
[B^\dagger(x,y)B(x,y)]^2\delta^4(x-y).  
\ebo
We now rescale the auxiliary field $B_i $ and the coupling $g_0$:
\bbo
B'_i= \sqrt{R} B_i(x,y)\qquad  \bar{g}_0=R^{-1/2} g_0
\qquad \text{where,} \qquad R= \bl 1 -\frac{g_0^2 N_c}{8 \pi^2 }\br. 
\ebo
We then have,
\bbo
\label{result0}
\backo\backo
S
=S_{KTH} + \!\int_x\bl [\bar{\psi}_Ri\slash{D}\psi_R]_f+[\bar{\psi}_Li\slash{D}\psi_L]_f  \br
+ \!\int_{xy} \!\bl \bar{g}_0 B'_i{}^\dagger (x,y)D_F(x-y)C^i(x,y)+h.c.\br
\nonumbo
-\!\int_{xy} \!
[B'^\dagger(x,y) D_F(x-y)B'(x,y)]
-\frac{\bar{g}_0^4\hat{\lambda}}{2M_0^6}\int_{xy}\!
D_F(x-y)[B'^\dagger(x,y)B'(x,y)]^2 +...,  
\ebo
where in the last term we use $\delta^4(x-y)\approx M_0^2D_F(x-y)$.


The ``equation of motion'' of the auxiliary field, $B'$,  using
eq.\ref{qc3}, then takes the form,
\bbo
\label{BBB}
B'_i(x,y)=\bar{g}_0C_{i}(x,y)
- \bar{g}_0^4 \hat\lambda M_0^{-6}B'_i(x,y) (B'{}^{j}{}^{\dagger}(x,y)B'_j(x,y)) +...,
\ebo
where we defined $C_i$ in eq.(\ref{C}).
We then solve eq.(\ref{BBB}) perturbatively in $\hat\lambda$
(where $\hat\lambda$ is defined in eq.(\ref{hatlam}):
\bbo
\label{tosolve}
 B'_i(x,y)\approx \bar{g}_0 C_{i}-
\bar{g}^7_0\hat\lambda M_0^{-6} C_{i} (
C^{\dagger j}(x)C_{j}(x) )+... 
\ebo
Substituting  into eq.(\ref{result0}), and noting cancellations
(particularly leading to the quartic term),
the action is then:
\bbo
\label{resultbb}
S= S_{KTH}+
\!\int_x\! \bl  
 [\bar{\psi}_L(x)i\slash{D}\psi_{L}(x)]+ [\bar{\psi}_R(x)\;i\slash{D}\psi_{R}(x)]\br
 \nonumbo +\bar{g}_0^2\!\int_{xy}
  \bl  N_cM_0^4H^\dagger(x,y) D_F(x-y)H(x,y)  
 +\;\sqrt{N_c}M_0^2(H^{i\dagger}(x,y)D_F(x-y)[\bar{\psi}_{R}(y)\psi_{iL}(x)]_f+h.c.)
 \nonumbo
- [\bar{\psi}^i_L(x)\psi_{R}(y)]_fD_F(x-y)[\bar{\psi}_{R}(y)\psi_{iL}(x)]_f\br
 - \frac{1}{2} \bar{g}^8_0 N_c\hat{\lambda} \int_{xy}\!\delta^4(x-y)
[H^{\dagger i}(x,y)H_i(x,y)]^2 + ...
\ebo
Therefore, we see that,
having integrated out the auxiliary field, we have obtained
an enhanced coupling in the $0^+$ binding channel:
 \bbo
 \label{r03}
\bar{g}_0^2= {g_0^2}\left(1- \frac{N_c g_0^2}{8\pi^2}\right)^{-1}.
\ebo
The rescaling of $g_0$ summed the  tower of fermion loops $\propto (g_0^2 N_c/8\pi^2)^n $,
which is seen by expanding $R^{-1}$.
The renormalized coupling $\bar{g}_0$ applies
to the binding action, and not to the interaction
of free  fermions with the colorons (e.g., for the decay width
of a coloron we would use the unrenormalized $g_0$).
On energy scales $<\!\!< M_0$ the enhanced $\bar{g}_0$ applies
in the 4-fermion interaction of free fermions, as is indicated in the effective action
in eq.(\ref{resultbb}). However, for energy scales $>\!\!> M_0$
the theory reverts to a pure topcolor gauge theory with coupling $g_0$
and the binding effects disappear.  
We see that $\bar{g}_0 > g_0$, so the underlying topcolor theory
is weaker than the effective binding  interaction. This is an example of a ``critical amplification'' effect
in the binding theory.

Passing to barycentric coordinates, $ H_i(x,y) \rightarrow \sqrt{2/J}\; H_i(X)\phi(r),$
and, $M_0^2D_F(2r) \rightarrow \delta^4(2r)$, we see that the Yukawa interaction becomes:
\bbo
\backo\backo
\bar{g}_0^2\!\int_{xy} \!\!\!\!
 \sqrt{N_c}M_0^2(H^{i\dagger}(x,y)D_F(x-y)[\bar{\psi}_{R}(y)\psi_{iL}(x)]_f+h.c.)
 \approx \bar{g}_Y\int d^4X \bl H^{i\dagger}(X)[\bar{\psi}_{R}(X)\psi_{iL}(X)]_f+h.c.\br.
\ebo
Therefore, we now have a renormalized Yukawa coupling: 
\bbo
\label{rgy}
\bar{g}_Y=  \bar{g}_0^2\sqrt{2N_c/J}\;\phi(0) .
\ebo
The physical top-Yukawa coupling is now given by $\bar{g}_Y$
and this will be set to its experimental value of unity as an input to define the theory
and obtain $M_0$.

Furthermore, 
we  have for the  quartic term  (where $\hat{\lambda}$ is defined in eq.(\ref{hatlam}): 
\bbo
\backo\backo
- \frac{1}{2} \bar{g}^8_0 N_c\hat{\lambda}\int_{xy}\!\delta^4(x-y)
[H^{\dagger i}(x,y)H_i(x,y)]^2
\longrightarrow -\frac{1}{2}\hat\lambda [\bar{g}^8_0  JJ^{-1} (2/J)^2N_c^2|\phi(0)|^4 ] \int_{X}(H^{i\dagger}(X) H_i(X))^2
\nonumbo
\equiv -\frac{\lambda}{2} \int_{X}\bl H^\dagger(X) H(X)\br^2
\qquad \text{where, using eq.(\ref{rgy}),} \qquad \lambda=  \frac{\bar{g}_Y^4 N_c}{4\pi^2}\ln\bl\frac{M_0}{m}\br.
\ebo
We can therefore write the leading terms in the full theory:
\bbo
\label{resultbb2}
S=
\!\int_x\! \bl  
 [\bar{\psi}_L(x)i\slash{D}\psi_{L}(x)]_f+ [\bar{\psi}_R(x)\;i\slash{D}\psi_{R}(x)]_f
 \br
 \nonumbo
 +\!\int_{X} \bl D_\mu H^\dagger(X) D^\mu H(X) +  H^\dagger(X) H(X) \int_r \bl -|\nabla_{\vvr} \phi(\vvr)|^2 +
 \left( \bar{g}_0^2  D_F(2r)|\phi(r)|^2 \right)\br
 \nonumbo
 \qquad + \bar{g}_Y(H^{i\dagger}(X)[\bar{\psi}_{R}(X)\psi_{iL}(X)]_f+h.c.)-\frac{\lambda}{2}  \int_{X}(H^\dagger(X) H(X))^2\br
 \nonumbo
 +\;\bar{g}_0^2\!\int_{xy} \; [\bar{\psi}^i_L(x)\psi_{R}(y)]_f D_F(x-y)[\bar{\psi}_{R}(y)\psi_{iL}(x)]_f+...
\ebo
With a solution to the SKG equation for $\phi(r)$ with eigenvalue $\mu^2$
(and relabeling the now dummy integration variable $X\rightarrow x$) we have: 
\bbo
\label{resultbb3}
S=
\!\int_x\! \bl  
 [\bar{\psi}_L(x)i\slash{D}\psi_{L}(x)]_f+ [\bar{\psi}_R(x)\;i\slash{D}\psi_{R}(x)]_f
 \br
 \nonumbo
 +\!\int_{x} \bl D_\mu H^\dagger(x) D^\mu H(x) -\mu^2 H^\dagger(x) H(x)  +\;\bar{g}_Y(H^{i\dagger}(x)[\bar{\psi}_{R}(x)\psi_{iL}(x)]_f+h.c.)
  - \frac{\lambda}{2} (H^\dagger(x) H(x))^2
 \br
 \nonumbo
 +\;\bar{g}_0^2\!\int_{xy} \; [\bar{\psi}^i_L(x)\psi_{R}(y)]_f D_F(x-y)[\bar{\psi}_{R}(y)\psi_{iL}(x)]_f
+ ...
\ebo
where $\mu^2<0$ for supercritical $\bar{g}_0^2$.
We have recovered the Standard Model BEH action with couplings to third generation $(t,b)$ quarks and the novel 4-fermion interaction
associated with coloron exchange. The covariant derivative, $D_\mu$, is that of the BEH boson eq.(\ref{covD}) (viewed either
as a low energy limit or explicitly with the incorporation
of Wilson lines as discussed in \cite{main}).

While we have extrapolated the loops to zero $P^2$, we
note that eq.(\ref{resultb}) contains a  $P^2$ term. If we interpret this
in the Wilsonian description of the NJL model, as in  \cite{main,BHL}, then eq.(\ref{resultb}) 
implies that the theory
is creating a loop level resonance  of the $B$ field with an action (ignoring gauge fields):
\bbo
\label{res}
\sim  \widetilde{Z} \partial_X B'^{i\dagger} \partial_X B'_i  -M_0^2 B'^{i\dagger} B'_i
\qquad \text{where,} \qquad
\widetilde{Z} = \frac{\bar{g}_0^2 N_c}{8 \pi^2 } \ln\left(\frac{M_0}{m}\right), 
\ebo
where $m\sim |P|$. The on-shell resonance mass near critical coupling is therefore
$
M_{resonance}^2 = M_0^2/  \ln\left(\frac{M_0}{m}\right) .
$
(Here we have reverted to interpreting $M_0^2 D_F \rightarrow \delta^4(2r)$).
Then eq.(\ref{tosolve}) is satisfied as $P^2<\!\!<M_0^2$.
The formation and exchange of this resonance below threshold $P^2\sim 0$ 
generates an attractive interaction
and can be viewed as leading to the enhancement of the effective coupling $\bar{g}_0^2$
as $P^2\rightarrow 0$.

\section{Discussion and Conclusions}

 \subsection{Physical Consequences of the Improved Yukawa Coupling}
 
 The renormalized topcolor coupling in the $0^+$ binding channel satisfies:
 \bbo
 \label{r0}
\bar{g}_0^2= {g_0^2}\bl 1- \frac{N_c g_0^2}{8\pi^2}\br^{-1}.
\ebo
where $g_0^2$ is the underlying topcolor coupling.
This represents ``critical amplification'' of the underlying coupling $g_0^2$
and all of the induced couplings, $g_Y$ and $\lambda$ (etc.).
It is $ \bar{g}_0^2 $ that controls
the formation of the semiclassical bound state.  This amplification effect
is not unexpected, since the quantum fermion loops
are essentially NJL model effects, and these are strongly binding.
We see that this can be understood as the
formation of a resonance  in the $0^+$ as
an NJL bound state, enhancing the effective coupling $\bar{g}_0^2$
as $P^2\rightarrow 0$.

The bound state is critical (has zero mass $\mu^2$) when $ \bar{g}_0^2 =  \bar{g}_{0c}^2 $ 
where we determine numerically \cite{main}:
 \bbo
\label{exact}
\frac{\bar{g}_{0c}^2N_c}{8\pi^2} =1.06940, \qquad \text{cf. NJL model:}\;\; 
\frac{\bar{g}_{0c}^2N_c}{8\pi^2}=1.
\ebo
This is the criticality condition of the Schr\"odinger equation with a Yukawa potential 
as translated into the SKG equation (note that we've often approximated $g_{0c}^2 \approx 8\pi^2/N_c$ 
using the NJL critical value in \cite{main}).

Solutions to the SKG equation lead to the eigenvalue $\mu^2$ which can be identified with
the SM symmetric phase, $\mu^2 = -(88)^2$ GeV$^2$, which is negative (tachyonic) 
for supercritical $\bar{g}_0^2> \bar{g}_{0c}^2$.
The solutions are discussed in \cite{main}.  For example, if we assume a 
large distance solution of the
form (the ``skeletal solution'' of Section IV.E of \cite{main} ),
then
 \bbo
 \phi(r) \approx  \frac{c e^{-|\mu| r}}{r}; \;\;\;\;\;\makebox{then we find,}\;\;\;\;\;
\phi(0)=\frac{2\sqrt{2}}{\pi^{3/2}} \sqrt{ \frac{|\mu|}{M_0}} =0.50795\sqrt{ \frac{|\mu|}{M_0}}.
 \ebo
 We can then fit the renormalized Yukawa coupling to experiment, 
 \bbo
 \backo\backo\backo
 \bar{g}_Y= 1 =\bar{g}^2_{0c}\sqrt{\frac{2N_c}{J}}\;\phi(0)=17.236\;\phi(0)
 \approx 8.7548 \sqrt{\frac{|\mu|}{M_0}},
 \ebo
 we then predict $M_0 = 6.745\; \makebox{TeV} $.\footnote{This was computed in \cite{main}
 where we quoted $5.9$ TeV, which resulted from using the NJL critical value $N_cg_{0c}^2/8\pi^2 =1$
 rather than the true $N_cg_{0c}^2/8\pi^2 =1.0694$. However, we think these are within
 the uncertainty of the leading order calculations for the SKG solutions, and does not
 take RG running effects, e.g., logarithms, into account.}
Likewise, using an approximate numerical solution to the SKG equation yields $M_0 = 5.23\; \makebox{TeV} $
(this is sensitive to a fitting function, and likely underestimates $M_0$).
Conservatively we would estimate $M_0$ to lie in a range of order $5 < M_0 < 7$ TeV. 

It should be emphasized that the collider physics predictions for properties of colorons
will depend mainly upon $g_0^2$, {\em not $\bar{g}_0^2$}.  The colorons form a QCD color octet, but with
the exception of couplings to $\bar{t}t$ would not be singly produced and would 
be pair produced in gluon fusion processes, which is sensitive to the QCD coupling and $M_0$.
Such coloron pairs lie near the energy limit of the LHC and will decay 
to pairs of $\bar{t}t$, and we also expect $\bar{b}b$ (topcolor must be extended to include the $b_R$-quark
and a $Z'$ to insure the $\bar{b}b$ channel is subcritical \cite{Topcolor}).
Coloron decay widths are controlled by $g_0^2$. If $\bar{g}_0^2$ is approximately critical,
$N_c \bar{g}_0^2/8\pi^2 \approx (1.0694)$
then from eq.(\ref{r0}) we have, 
\bbo
\frac{g_0^2 N_c }{8\pi^2}\approx \frac{1.0694}{1+1.0694}=0.51677,
\ebo
and the underlying topcolor theory is subcritical, and  $\alpha_{topcolor} = \frac{g_0^2  }{4\pi}\approx 1.0823 $
is marginally perturbative.

 \subsection{Physical Consequences of the Improved Quartic Coupling}
 
For the quartic coupling
we obtain above:  
\bbo
\label{quartic1}
\backo
{\lambda}\approx \frac{\bar{g}_Y^{4}N_c}{4\pi ^{2}}%
 \ln \left(\frac{M_0}{\mu }\right).
\ebo
The log evolution is just that of a single fermion loop 
and matches the result for the pointlike NJL case, with $g_{top}=\bar{g}_Y$.
Not surprisingly when we take the pointlike potential limit the loop
result of the bilocal theory confirms a pointlike NJL calculation
(see discussion of Section 4 of ref.(\cite{main2}). 

Experimentally, in the SM using the value of $m_{BEH}\approx 125$ GeV and $v_{weak}\approx 175$ GeV, 
we find $\lambda \approx 0.25$.
In  the old pointlike NJL top condensation model the quartic coupling
was determined by the RG with ``compositeness boundary conditions,''
where we obtained (running-down from the Landau pole at $M_0$ to $|\mu|$), the result $\lambda\sim 1$.
This is too large and leads to predicted $m_{BEH}\sim 260$ GeV. Indeed, the quartic coupling is generally problematic for NJL based theories of a composite BEH boson.

However, in the present bilocal scheme owing to suppression of $\bar{g}_Y\propto \phi(0)$, the quartic coupling
is also suppressed $\propto |\phi(0)|^4$ and is now generated in RG running, from a value
of zero at $M_0=6$ TeV, down  to $|\mu|\sim 88$ GeV, using the induced
$\bar{g}_Y$.  Keeping only the $\bar{g}_Y^4$ contribution
we would obtain numerically from eq.(\ref{quartic1}):
\bbo
\lambda \approx (\bar{g}_Y^{4}) \frac{N_{c}}{4\pi ^{2}}%
 \ln \left( \frac{M_0}{|\mu| }\right) \approx 0.321\;\;\; \makebox{(cf., $0.25 $ experiment).}
\ebo
This result is significantly better than the old NJL top condensation model.
However,  prefactor at one loop level should reflect the full RG running of $\lambda$,
(see, e.g., \cite{HillTA}), which would yield at leading log approximation:
\bbo
\lambda \approx (\bar{g}_Y^{4}-\bar{g}_Y^2\lambda-\lambda^2) \frac{N_{c}}{4\pi ^{2}}%
 \ln \left( \frac{M_0}{|\mu| }\right) \approx 0.230\;\;\; \makebox{(cf., $0.25 $ experiment).}
\ebo  
This includes the $\lambda^2$ term which arises from loops with propagating internal BEH boson.
For the present composite model this not likely to be correct, due to the extended size of the BEH wave-function.
The terms of order $\bar{g}_Y^2\lambda$ are leg-renormalizations, due to the top quark loop and presumably can
be retained. This would imply a result:
\bbo
\lambda \approx (\bar{g}_Y^{4}-\bar{g}_Y^2\lambda) \frac{N_{c}}{4\pi ^{2}}%
 \ln \left( \frac{M_0}{\mu }\right) \approx 0.243\;\;\; \makebox{(cf., $0.25 $ experiment).}
\ebo

\subsection{ Summary}

We have used the techniques of Jackiw,  {\etal}$\!\!,$  \cite{Jackiw,CJT} to 
obtain formally the effective action for the semiclassical third generation fields $\psi_L,\psi_R$
in a single coloron exchange interaction field theory of topcolor.
This describes the Brout-Englert-Higgs boson as a composite object with constituents that are
the semiclassical fields composed of $t$ and $b$ quarks.

This marks a radical departure from the old RG improved NJL
model which builds a composite state by integrating out all of the fermions, 
in analogy to confinement in QCD.  For us, the bound state is softer, and more
Hydrogenic, containing a subset of the low momentum modes in the semiclassical theory.
This gives us the internal wave-function $\phi(r)$. Tuning the theory within
a few percent of the critical coupling yields a hierarchy that defines a low mass
BEH bound state. We have given the central results in the companion paper \cite{main}
and addressed the most significant quantum effects presently.
Here we have introduced  an auxiliary field, $B_i$, which consolidates the main large-$N_c$ fermion
loop contributions. Upon computing these, we integrate out $B_i$ to obtain
the full quantum corrected action.

Our most significant new result is that the topcolor coupling, $g_0^2$ is subcritical while the 
effective coupling in the binding potential $\bar{g}_0^2$ is enhanced and can be slightly supercritical:
\bbo
\bar{g}_0^2= {g_0^2}\bl{1- \frac{N_c g_0^2}{8\pi^2}}\br^{-1}
\ebo
This in turn renormalizes the Yukawa coupling and quartic coupling. 
Hence the value of $\bar{g}_0^2$ is ``critically amplified'' above a weaker
value of the underlying topcolor coupling $g_0^2$.

Inputting the experimental result for the
renormalized Yukawa coupling, $\bar{g}_Y=1$ with the symmetric
phase BEH mass, $\mu^2=-(88)^2$ GeV$^2$, yields the 
central prediction obtained for $M_0\approx 6$ TeV as discussed in \cite{main}. 
Hence the major results of \cite{main} still hold, but with $g_0^2$ now replaced by $\bar{g}_0^2$.
 Due to the linear relationship between $\bar{g}_0^2$ and $|\mu|/M_0$, a consequence
 of the dilution effect of $\phi(0)$ as discussed in \cite{main}, 
we see the degree of fine-tuning of the hierarchy is of order
$\sim |\mu|/M_0 \sim 1.4\%$.

We explicitly demonstrate that the BEH boson quartic coupling, $\lambda$, arises from loops 
$\sim N_c \bar{g}_Y^4\ln(M_0/\mu)/4\pi^2 \propto (\phi(0))^4$.
The result for the quartic coupling is $\lambda\sim 0.24$, compared
to $\lambda\approx 0.25$ experimentally.
The $\lambda$ prediction is in astonishingly good agreement with experiment and we believe makes 
the natural top condensation idea compelling.

The ``colorons'' which mediate the binding interaction form a 
QCD color octet.
They have a global $SU(3)$ symmetry and a conserved $SU(3)$ current, and must 
therefore be pair-produced.  They will decay in the minimal model to $\bar{t}t$,
but in more realistic topcolor models 
they can also decay to $\bar{b}{b}$ (\cite{Topcolor}\cite{Topcolor2}\cite{NSD} and references therein).
They may be accessible
 to the LHC in the multi-TeV range \cite{Burdman,expt,limits}, favoring the third generation
in its couplings.  It should be straightforward to obtain lower bounds on $M_0$ from single gluon or gluon fusion
 production of pairs of the colorons.  
 The colorons will produce excesses in 4-top, $t\bar{t}t\bar{t}$, events.  In
 an extended model we would also expect $t\bar{t}b\bar{b}$ and $b\bar{b}b\bar{b}$ anomalies
 to emerge.   The more general topcolor models offer many possibilities
 within the arena of heavy quark flavor physics.  

This theory, if confirmed, solves the ``naturalness problem'' of the BEH boson in the Standard Model
and opens up a vista of new physics with new gauge interactions.
Many avenues for further theoretical development exist.  Notably, a revisitation 
of the Topcolor $Z'$ \cite{Topcolor}, resonances, and new flavor physics
associated with $\bar{b}b$ would be of interest.
There is much to do to further develop these
models in the context of the present dynamics.

\vspace{0.1in}

\section*{Acknowledgments}

I thank Bill Bardeen, Thom Curtright, Bogdan Dobrescu,  Andreas Kronfeld 
and Julius Kuti for discussions and comments during the course of this work.

\end{document}